\documentclass[runningheads]{llncs}
\usepackage{graphicx}
\usepackage{times}  
\usepackage{helvet} 
\usepackage{courier}  
\usepackage[hyphens]{url}  
\usepackage{xcolor}
\usepackage[ruled,vlined]{algorithm2e}
\begin{document}
\title{Characterizing Sociolinguistic Variation in the Competing Vaccination Communities}
%
%
\author{Shahan A. Memon\inst{1}\orcidID{0000-0002-1152-0867} \and Aman Tyagi\inst{2}\orcidID{0000-0002-6654-0670} \and David R. Mortensen\inst{1}\orcidID{0000-0002-9096-6137} \and Kathleen M. Carley\inst{1,2}\orcidID{0000-0002-6356-0238}}


%
\authorrunning{SA. Memon et al.}
%
\institute{School of Computer Science , Carnegie Mellon University, PA 15213, USA \and 
Engineering and Public Policy, Carnegie Mellon University, PA 15213, USA
\email{samemon@cs.cmu.edu,amantyagi@cmu.edu, \{dmortens,kathleen.carley\}@cs.cmu.edu}}
\maketitle              
\begin{abstract}
Public health practitioners and policy makers grapple with the challenge of devising effective message-based interventions for debunking public health misinformation in cyber communities. \emph{Framing} and \emph{personalization} of the message is one of  the key features for devising a persuasive messaging strategy.   For an effective health communication, it is imperative to focus on \emph{preference-based framing} where the preferences of the target sub-community are taken into consideration. To achieve that, it is important to understand and hence characterize the target sub-communities in terms of their social interactions. In the context of health-related misinformation, vaccination remains to be the most prevalent topic of discord. Hence, in this paper, we conduct a sociolinguistic analysis of the two competing vaccination communities on Twitter: \emph{pro-vaxxers} or individuals who believe in the effectiveness of vaccinations, and \emph{anti-vaxxers} or individuals who are opposed to vaccinations. Our data analysis show significant linguistic variation between the two communities in terms of their usage of linguistic intensifiers, pronouns, and uncertainty words. Our network-level analysis show significant differences between the two communities in terms of their network density, echo-chamberness, and the EI index. We hypothesize that these sociolinguistic differences can be used as proxies to characterize and understand these communities to devise better message interventions.
\end{abstract}
\section{Introduction}

Health-related misinformation has detrimental effects on the public health. According to researchers, many preventable diseases have re-emerged as a consequence of the drop in immunization rates due to declining trust in vaccines caused by the misinformation on the web \cite{usnews2}. Moreover, distrust in vaccines and health expertise is only expected to increase in the next decade \cite{johnson2020online}. 

Debunking public health misinformation requires an effective health communication such as a \emph{message-based intervention}. For an effective message-based intervention, it is imperative to focus on \emph{preference-based framing} where the preferences of the target sub-community are taken into consideration. These preferences can be defined over three main aspects: (i) choice of the messenger, (ii) medium of information dissemination, and (iii) content of the message. A message intervention is effective if the message is delivered by a \emph{trusted} source using an optimal medium of dissemination. In online communities, this translates to identifying the \emph{influencers} or nodes with high \emph{degree centrality} in the social network such as shown in \cite{sanawi2017vaccination}. Choosing the content of the message, on the other hand, requires a thorough understanding of how the target community members interact with each other, what \emph{language choices} they make, and how those language choices reflect their \emph{non-negotiable social identities}. 

Vaccination related misinformation is arguably the most prevalent form of misinformation online. Therefore, for the purposes of this study, we chose to tap into vaccination discourse on Twitter. We study the conversations between two competing groups of Twitter users: (i) those who believe in the effectiveness of vaccinations \emph{(pro-vaxxers)}, and (ii) those who are skeptical \emph{(anti-vaxxers)}. The goal of our study is to characterize the two competing vaccination communities in terms of their sociolinguistic variation. We hypothesize that understanding the interactions of the members of these communities can help devise a better messaging strategy.

Prior work includes the sociolinguistic analysis of Twitter in multilingual societies \cite{kim2014sociolinguistic}, predicting community membership using word frequencies \cite{bryden2013word}, identifying effective vaccine communication using fuzzy trace theory \cite{broniatowski2016effective}, understanding the evolution of competing views around vaccination at the system level \cite{johnson2020online}, and sociolinguistic study of online echo-chambers \cite{duseja2019sociolinguistic}. 

We extend the work by Duseja and Jhamtani in \cite{duseja2019sociolinguistic} to study vaccination-based communities on Twitter by understanding their differences in usage of linguistic intensifiers, pronouns, and uncertainty words. We also conduct a network-level analysis by computing the network density, EI index, and echo-chamberness for the two target communities.

\section{Dataset}
To construct our dataset, we employ a three-stage process: (i) we first collect data using a set of hashtags via the Twitter search and the Twitter streaming API; (ii) we use this data to identify the two communities; and (iii) finally, to mitigate survivorship bias \cite{brown1992survivorship} and collect more data per individual, we collect timelines of the identified pro- and anti-vaxxers. We describe this process in detail in the following subsections. In the section \ref{datastats}, we present the statistics for the final set of data we use to conduct our analyses.

\subsection{Data Collection}
\label{datacollection}
We first collect a set of known pro-vaccination and anti-vaccination hashtags from our domain knowledge as well as from the background literature \cite{dredze2017vaccine}. List of these hashtags can be found in Table \ref{tab:hashtags}.
We use these hashtags to collect Twitter data through the Twitter Streaming API, and augment it with data collected from Twitter Search API. The data consists of Tweets from 29th October 2019 to 12th November 2019. Based on \cite{broniatowski2018weaponized}, we filter out all tweets that do not include the lemmas ``vacc" or ``vax" (case insensitive) as part of their tweet text. This is to remove any possible noise in the data.

\begin{table}[!htb]
\caption{This table shows the hashtags used for the task of data collection. We use camel-casing for better readability.}
\resizebox{\columnwidth}{!}{
\begin{tabular}{|c|c|} 
\hline
Stance & Hashtags \\
\hline
\hline
Pro-vaccination  & \begin{minipage}{30em}\textit{VaccinesSaveLives, VaccinesWork, WorldImmunizationWeek, VaxWithMe, HealthForAll, WiW, ThankYouLaura}\end{minipage}\\
\hline
Anti-vaccination & \begin{minipage}{30em}\textit{LearnTheRisk, VaccineInjury, VaccineDeath, VaccineDamage, VaccinesCauseAutism, CDCFraud, CDCWhistleBlower, CDCTruth, WakeUpAmerica, HearUs, HealthFreedom}\end{minipage}\\
\hline
Unidentified & \begin{minipage}{30em}\textit{Vaccine, Vaccines, Vaccinate, VaccinateUS}\end{minipage}\\
\hline
\end{tabular}
}
\vspace{1em}
\label{tab:hashtags}
\end{table}
\subsection{Community Detection}
\label{communitydetection}

\subsubsection{Label Propagation}
To be able to conduct any analysis, it is imperative to identify the competing groups. Assigning a stance to a tweet or a twitter user is a non-trivial problem. Therefore, we use a similar method as described in \cite{tyagi2020brims,tyagi2020computational} to find anti-vaxxers and pro-vaxxer groups based on the weighted combination of the \emph{valence} of their hashtags. In this study, we assume that retweets indicate endorsement.

In the previous studies such as \cite{evans2016stance}, hashtags have been shown to work as realistic proxies for identifying stances among different groups on social media sites. In \cite{tyagi2020brims}, hashtags are used to identify twitter users who believe in anthropogenic causes of climate change and those who do not. Similarly, in \cite{tyagi2020computational}, hashtags could also be used to identify polarization in political discourse and how the polarization can change with time.

We use community detection method based on the work done in \cite{tyagi2020computational}. We first choose 2 seed hashtags for each of the polarized groups: \textit{\#VaccinesSaveLives} and \textit{\#VaccinesWork} for pro-vaccination and \textit{\#VaccineInjury} and \textit{\#LearnTheRisk} for anti-vaccination. We assign pro-vaccination seeds a valence of +1, and anti-vaccination seeds a valence of -1. \footnote{We randomly sample 100 tweets for each of these hashtags. For pro-vaccination hashtags, $98\%$ of tweets with hashtag \textit{\#VaccinesSaveLives} and $97\%$ of tweets with hashtag \textit{\#VaccinesWork} were related to pro-vaccination. For anti-vaccination hashtags, $88\%$ of tweets with hashtag \textit{\#LearnTheRisk} and $93\%$ of tweets with hashtag \textit{\#VaccineInjury} were related to anti-vaccination.} We then create a hashtag co-occurrence graph to identify most co-occurring hashtags with the chosen seeds, and choose those that are semantically similar, as well as the ones that are known to be pro-vax and anti-vax hashtags from the background literature \cite{dredze2017vaccine,broniatowski2018weaponized,broniatowski2016effective} to manually assign a hard valence of +1 and -1. We then use a variant of label propagation algorithm \cite{xiaojin2002learning} described as Algorithm \ref{alg:labelprop} below to assign valence to each of the remaining hashtags. Similar to \cite{tyagi2020computational} the input to the algorithm is a hashtag-to-hashtag co-occurrence graph where hashtags represent nodes, and nodes are connected if they co-occur. The edges are weighted by the frequency of co-occurrence.

\begin{algorithm}[h]
\small
\SetAlgoLined
\KwIn{Nodes = \emph{n}; Edges = \emph{e}; Edge Weight = \emph{$e_{ij}$}, $i \in n$ and $j \in n$}
 \textbf{initialize} $\gamma=50$ and \emph{i}\;
 \For{each n}{
  \textbf{define} $l = integer(i/\gamma)$; $i+=1$\;
  \For{each n}{
      \If{n not labeled}{
       \textbf{compute} $t$ = neighbors of $n$\;
       \textbf{compute} $t_l$ = labeled neighbors of $n$\;
       \If{$|t_l| + l \geq t$}{
        \textbf{initialize} \textit{score}, $c$\\
        \For{each $t_i \in t$}{
            score += label $t_i$ * $e_{n{t_i}}$\\
            c += $e_{n{t_i}}$\\
            }
        \textbf{update} label $n = score/c$
       }
        }
    }
 }
\caption{Label Propagation Algorithm}
\label{alg:labelprop}
\end{algorithm}

\subsubsection{Stance Identification}
Once we have identified the valence of a set of hashtags, we aggregate hashtags used by each user and find a weighted average of the valence of all hashtags used by a particular user. We label a user as pro-vaxxer, or anti-vaxxer if the weighted average was positive, or negative respectively.

Using the algorithm, 3295 users are identified as pro-vaxxers, 2967 as anti-vaxxers. We randomly sample 100 users that were classified as pro-vaxxers and 100 users that were classified as anti-vaxxers to evaluate the quality of assignment. We find $96\%$ of the labeled pro-vaxxers as pro-vaxxers, and $80\%$ of the labeled anti-vaxxers as anti-vaxxers. 

\subsection{Timeline Extraction}

Both Twitter streaming API and the Twitter search API do not allow the collection of data beyond a certain time period to be able to extract historical tweets. As a consequence, we collect our initial set of tweets within a fixed time window of 15 days. Because our goal was to study how the non-negotiable social identities of users correlated to their linguistic choices on Twitter, windowing the data by time period of 15 days could lead to high survivorship bias where users with higher activity within the chosen days could introduce bias in our analyses by having a higher influence. This is why, we decided to augment our data with timelines of identified individual users. This may not remove the survivorship bias completely, but may help mitigate it.

At the end of timeline extraction, we only retain one copy of each of the tweets. More concretely, to avoid over-inflating the effect of certain tweets that are more viral than the other, we use only unique tweet texts. This is an important preprocessing step to conduct a sociolinguistic frequency-based analysis.

\subsection{Data Statistics}
\label{datastats}

At the end, our sociolinguistic analysis is conducted on an overall 6262 Twitter users with an aggregate of 588,110 tweets. This included 3295 pro-vaxxers with 310461 pro-vaccination tweets, and 2967 anti-vaxxers with 277649 anti-vaccination tweets, making it an average of about 94 tweets per user for both pro- and anti-vaxxers. 

\section{Methodology}
We conduct two types of analyses to characterize the two competing groups: \emph{linguistic analysis} and \emph{network analysis}. 
\subsection{Linguistic Analysis}
We test three linguistic variables which are described as follows.

\subsubsection{Linguistic Intensification}
We first study the differences in the usage of linguistic intensifiers. Intensifiers are words, or phrases that strengthen the meaning of other expressions and show emphasis. Examples include amplifiers (eg.``really", ``very"), usage of swear words, general interjections (eg. ``wow", ``omg"), and exclamations. Intensifiers are commonly used to bolster argumentation to persuade the target audience. We hypothesize that users that are pro-vaxxers use more intensifiers. This is because pro-vaxxers have been found to frequently debunk anti-vaxxers' claims with scientific evidence \cite{boser2018mothers}. Therefore, they would seem to take the corrective approach intended to persuade anti-vaxxers, hence using more intensifiers.

\subsubsection{Pronominal Usage}
Pronouns play a key role in models of narrative and discourse processing \cite{gibbons2018pronouns}. Because most of the vaccine-related misinformation is based on personal anecdotes, we would expect pronominal usage to be high amongst anti-vaxxers. To test this, we identify various different categories of pronouns (eg.  ``subject pronouns", ``object pronouns", ``third-person pronouns"), a complete list of which can be found in Table \ref{tab:linguisticcats}.

\subsubsection{Use of Uncertainty Words}
Previous research \cite{duseja2019sociolinguistic}  has found the use of uncertainty words (eg. ``might", ``likely") as a negative linguistic correlate of echo-chamberness. This is based on the hypothesis that because users not in echo-chambers are exposed to alternate views, they may be less certain of their ideas. We adopted the list of uncertainty words from \cite{duseja2019sociolinguistic} to test if that is true i.e. if there is a significant difference in the use of uncertainty words across the two vaccination communities.

\begin{table}[!ht]
\scriptsize
\caption{This table shows the lexical categories we use for the sociolinguistic analysis along with the chosen list of words for each category (lexicon).}
\centering
\begin{tabular}{|c|c|} 
\hline
Lexical Category & Lexicon (vocabulary)\\
\hline
\hline
\textbf{Intensifiers} & \\
\hline
\hline
Amplifiers & \begin{minipage}{30em}
     \textit{amazingly, -ass, astoundingly, awful, bare, bloody, crazy, dead, dreadfully, colossally, especially, exceptionally, excessively, extremely, extraordinary, fantastically, frightfully, fucking, fully, hella, holy, incredibly, insanely, mad, mightily, moderately, most, outrageously, phenomenally, precious, quite, radically, rather, real, really, remarkably, ridiculously, right, sick, so, somewhat, strikingly, super,supremely, surpassingly, terribly, terrifically, too, totally, uncommonly, unusually, veritable, very, wicked}
    \end{minipage}\\
\hline
Swear words & \begin{minipage}{30em}\textit{fu*****, etc.} A complete list of words can be found on Wikipedia's English swear words page \cite{}. \end{minipage}\\
\hline
General interjections & \textit{wow, hooray, ouch, uh oh, ew, aw, omg}\\
\hline
Exclamation & \textit{!*}\\
\hline
\hline
\textbf{Uncertainty words} & \begin{minipage}{30em}
     \textit{may, might, perhaps, maybe/may-be, potentially, possibly, likely, probably, probable, possible, think, seem, believe, presume, would be, could be}
    \end{minipage}\\
\hline
\hline
\textbf{Pronouns} & \textit{}\\
\hline
Demonstrative & \textit{this, that, these, those}\\
\hline
Possessive & \textit{ours, mine, yours, theirs, his, hers}\\
\hline
Quantifier & \begin{minipage}{30em}\textit{few, several, some, all, much, one, fewer, many, more, most, plenty, less, little, enough}\end{minipage}\\
\hline
Reflexive & \begin{minipage}{30em}\textit{myself, herself, ourselves, themselves, yourself, himself, itself, yourselves}\end{minipage}\\
\hline
First-Person & \textit{I, we, us, me, myself, my, mine, our, ours}\\
\hline
Second-Person & \textit{you, yours, you’re, your}\\
\hline
Third-Person & \begin{minipage}{30em}\textit{he, she, theirs, themselves, them, her, him, his, himself, hers, herself, it, its, itself, they}\end{minipage}\\
\hline
Gendered third-person & \textit{he, she, her, him, his, himself, hers, herself}\\
\hline
Subject & \textit{I, she, he, they, we, you, it}\\
\hline
Object & \textit{me, us, them, him, you, her, it}\\
\hline
IT & \textit{it, it’s, its, itself}\\
\hline
\hline
\end{tabular}
\vspace{1em}
\label{tab:linguisticcats}
\end{table}

\subsection{Network Analysis}
We also compute three network-level measures to characterize the network structure of the two target communities. We describe each of these measures in detail in their respective sections below.

\subsubsection{Network Density}
Network density is defined as the ratio of actual connections and potential connections \cite{giuffre2015cultural}. Dense networks tend to ``groupthink" \cite{smelser2001international} where conformity of ideas is highly valued and difference of opinions is discouraged. 

\subsubsection{EI Index}
The EI (External-Internal) index was developed by Krackhardt and Stern in \cite{krackhardt1988informal} as a measure of dominance of external over internal ties. More concretely, assuming two groups based on some attribute, one group defined as internal and the other as external, the EI index is computed as follows:

\begin{equation}
EI= \frac{EL-IL}{EL+IL}
\label{eq:ei}
\end{equation}

where EL represents the number of external links and IL represents the number of internal links. EI index is a useful proxy for identifying echo-chamberness.

\subsubsection{Echo-chamberness}
To compare the echo-chamber effect in the two vaccination groups, we also directly compute the echo-chamberness of the two communities. We use the following definition of echo-chamberness: For a given network $G$, the echo-chamberness (EC) is defined as:

\begin{equation}
EC = (r * d)^{1/3}
\label{eq:ec}
\end{equation}

where r is the reciprocity \cite{wasserman1994social} of graph G or the ratio of bi-directional edges and the total number of edges in G, and d is the density of graph G. 

\subsection{Evaluation}
\subsubsection{Test Statistics} For each sub-category of the linguistic features in Table \ref{tab:linguisticcats}, we use two test statistics to compute the difference between the two groups. These are as follows: 
\begin{enumerate}
    \item The overall proportion of tweets that contain any of the words for a given lexical category ($T_1$)
    \item The mean of the proportions of tweets of individual users containing any of the words for a given lexical category ($T_2$)
\end{enumerate}
We use these test statistics to compute (i) the difference of proportions between the two groups, and (ii) the difference of means of proportions between the two groups.

The first test statistic regards each tweet independently. We use the second test statistic to account for differences in the linguistic choices of individual users.

\subsubsection{Statistical Significance:} For the first statistic, we use a two-sample z-test for the difference of proportions ($Z_1$). For the second statistic, we use an independent z-test for the difference in means ($Z_2$). For all the tests, our $\alpha = 0.05$.

\section{Results and Discussion}
\subsection{Linguistic Analysis}
The summary of our linguistic analysis across all the lexical categories can be found in Table \ref{tab:linganalysis}.

\begin{table}[!ht]
\scriptsize
\caption{This table shows the summary of our analyses across all the linguistic categories. The first column shows the lexical category. The second and third columns show the first test statistic as a percentage for pro-vaxxers and anti-vaxxers respectively. The fourth and fifth column display the z-score and p-value for the z-test for the difference of proportions. The sixth and seventh columns show the second test statistic as a mean percentage for pro-vaxxers and anti-vaxxers respectively. The eighth and ninth columns display the z-score and p-value for the independent z-test for the difference in means}
\centering
\begin{tabular}{|c||c|c|c|c||c|c|c|c|} 
\hline
Lexical Category & $T_1$ (Pro) & $T_1$ (Anti) & z-score ($Z_1$) & p-value ($Z_1$) &  $T_2$ (Pro) & $T_2$ (Anti) & z-score ($Z_2$) & p-value ($Z_2$)\\
\hline
\hline
\textbf{Intensifiers} & 45.90\% & 50.60\% & -36.25 & $<.001$ & 11.63\% & 14.96\% & -6.59 & $<.001$\\
\hline
Amplifiers & 31.40\% & 37.10\% & -45.32 & $<.001$ & 10.91\% & 13.66\% & -5.66 & $<.001$\\
\hline
Swear words & 4.0\% & 5.60\% & -27.40 & $<.001$ & .57\% & 1.04\% & -3.26 & $<.001$\\
\hline
General interjections & 17.50\% & 16.70\% & 7.89 & $<.001$ & .43\% & .58\% & -1.37 & .17\\
\hline
Exclamation & 1.10\% & 2.20\% & -34.17 & $<.001$ & - & - & - & -\\
\hline
\hline
\textbf{Uncertainty words} & 5.7\% & 7.0\% & -20.84 & $<.001$ & 4.12\% & 5.07\% & -3.23 & .001\\
\hline
\hline
\textbf{Pronouns} & 55.80\% & 62.20\% & -49.68 & $<.001$ & 55.94\% & 61.83\% & -7.38 & $<.001$\\
\hline
Demonstrative & 17.63\% & 20.91\% & -31.84 & $<.001$ & 18.61\% & 21.73\% & -5.20 & $<.001$\\
\hline
Posessive & 1.30\% & 1.60\% & -9.39 & $<.001$ & 1.49\% & 1.67\% & -.92 & .36\\
\hline
Quantifier & 15.3\% & 16.0\% & -6.70 & $<.001$ & 15.20\% & 16.83\% & -3.06 & .002\\
\hline
Reflexive & .80\% & .86\% & -2.26 & .02 & 1.49\% & .92\% & 3.43 & $<.001$\\
\hline
First-Person & 21.20\% & 23.44\% & -20.67 & $<.001$ & 20.96\% & 22.54\% & -2.45 & .01\\
\hline
Second-Person & 16.40\% & 18.5\% & -20.69 & $<.001$ & 15.22\% & 16.47\% & -2.23 & .03\\
\hline
Third-Person & 14.8\% & 20.9\% & -60.51 & $<.001$ & 14.29\% & 20.84\% & -11.74 & $<.001$\\
\hline
Gendered third-person & 3.60\% & 5.60\% & -36.84 & $<.001$ & 3.15\% & 4.92\% & -5.96 & $<.001$\\
\hline
Subject & 28.90\% & 37.50\% & -69.53 & $<.001$ & 27.64\% & 35.55\% & -10.89 & $<.001$\\
\hline
Object & 21.64\% & 26.90\% & -46.77 & $<.001$ & 19.66\% & 24.51\% & -7.91 & $<.001$\\
\hline
IT & 8.30\% & 10.29\% & -26.16 & $<.001$ & 8.21\% & 9.44\% & -3.07 & .002\\
\hline
\hline
\end{tabular}
\vspace{1em}
\label{tab:linganalysis}
\end{table}
\subsubsection{Linguistic Intensification}
We observe that our initial hypothesis that pro-vaxxers use more intensifiers is false. What we find is that anti-vaxxers employ significantly more linguistic intensifiers than pro-vaxxers. This holds true across all the sub-categories of intensifiers with the exception of the use of general interjections where the difference is marginal and not significant. While intensifiers are used as a persuasion technique, the observed results can possibly be explained by an old theory in speech communication that correlates the use of intensifiers with perceived powerlessness \cite{bradac1995men,hosman1989evaluative}. Intensifiers and hedges are used more generally by people with low social power \cite{bradac1995men}. Because anti-vaxxers are a minority group, it is a possible argument one could make as perceived  minority leads to perceived low social power which could lead to high linguistic intensification.
\subsubsection{Pronominal Usage}
From our analyses, we find that with the exception of reflexive and possessive pronouns, anti-vaxxers show a significantly high pronominal usage across all the categories. This difference is prominent specifically for third-person, gendered third-person, subject, and object pronouns. In sociolinguistic literature, pronouns are predominantly linked with narrative discourse structure. For example object pronouns such as ``him" or ``his" and gendered third-person pronouns ``he" or ``she" have a referential property, where their semantic interpretation is dependent on what they are referring to. Anaphoric references define objects already defined in the discourse \cite{young2004systemic} which creates a better narrative viewpoint. Like intensifiers, pronouns are also found to be used heavily by people with lower levels of perceived power \cite{nerbonne2014secret}. 

\subsubsection{Use of Uncertainty Words}
In terms of the use of uncertainty words, while we do find a significant difference between the two communities, we do not observe the same effect observed in the background literature \cite{duseja2019sociolinguistic}. In fact, we find a counter-intuitive result i.e. that the anti-vaccination community with higher echo-chamberness (as observed in section \ref{sec:networksec}) tends to use more uncertain words than pro-vaccination community. This is an evidence that not all echo-chamber communities show certainty in their tweets as observed in \cite{duseja2019sociolinguistic}.

\subsection{Network Analysis}
\label{sec:networksec}
\begin{figure}[!htp]
\caption{Mention (left), retweet (middle), and reply (right) networks of pro (in green) and anti (in red) vaccination communities created using ORA-PRO \cite{altman2018ora,carley2017ora}}
    \centering
    \includegraphics[width=4cm]{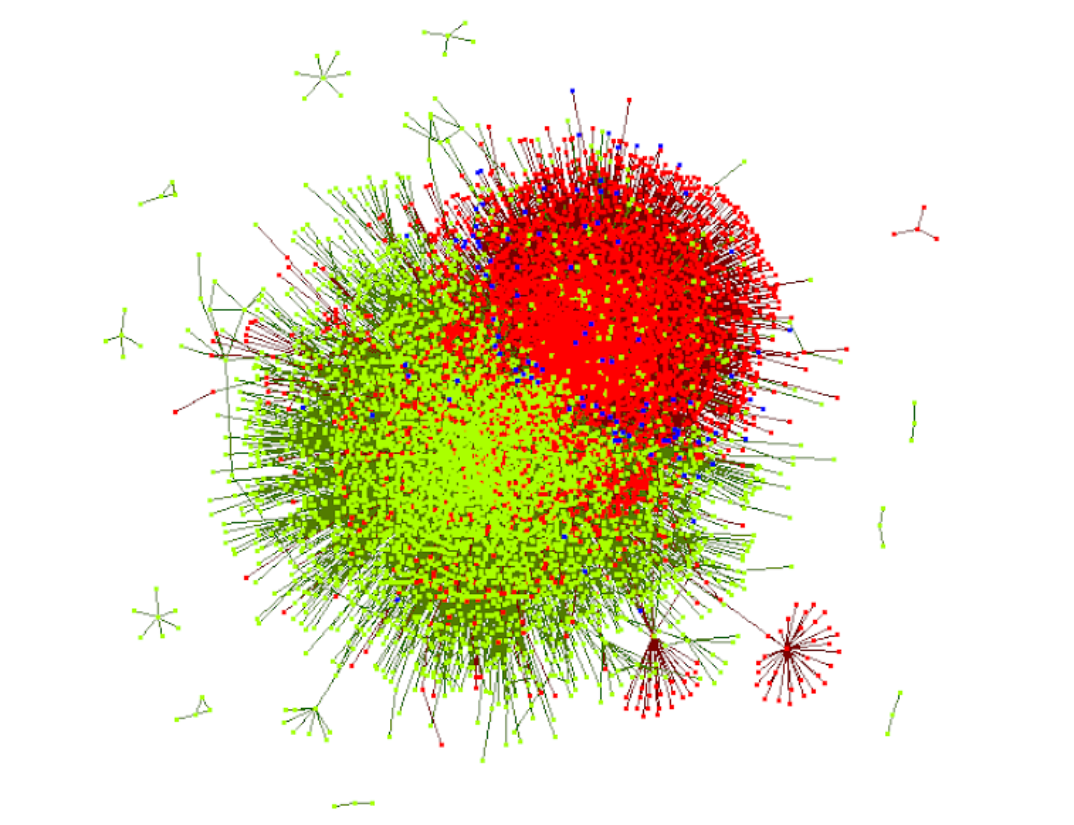}
    \includegraphics[width=3cm]{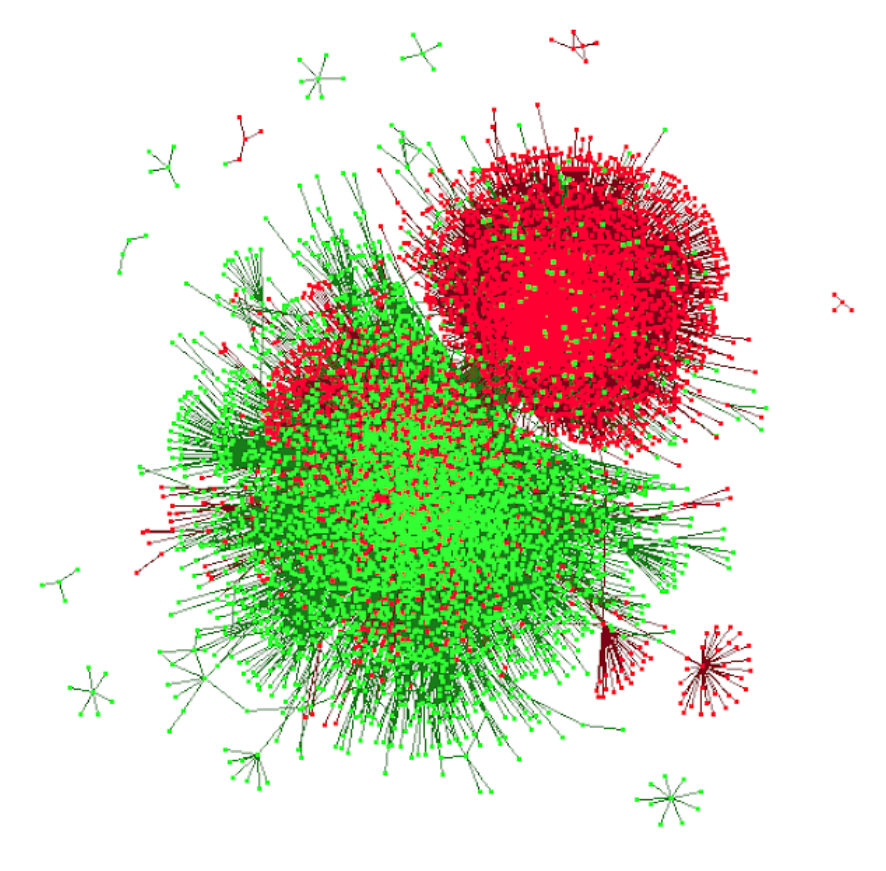}
    \includegraphics[width=3cm]{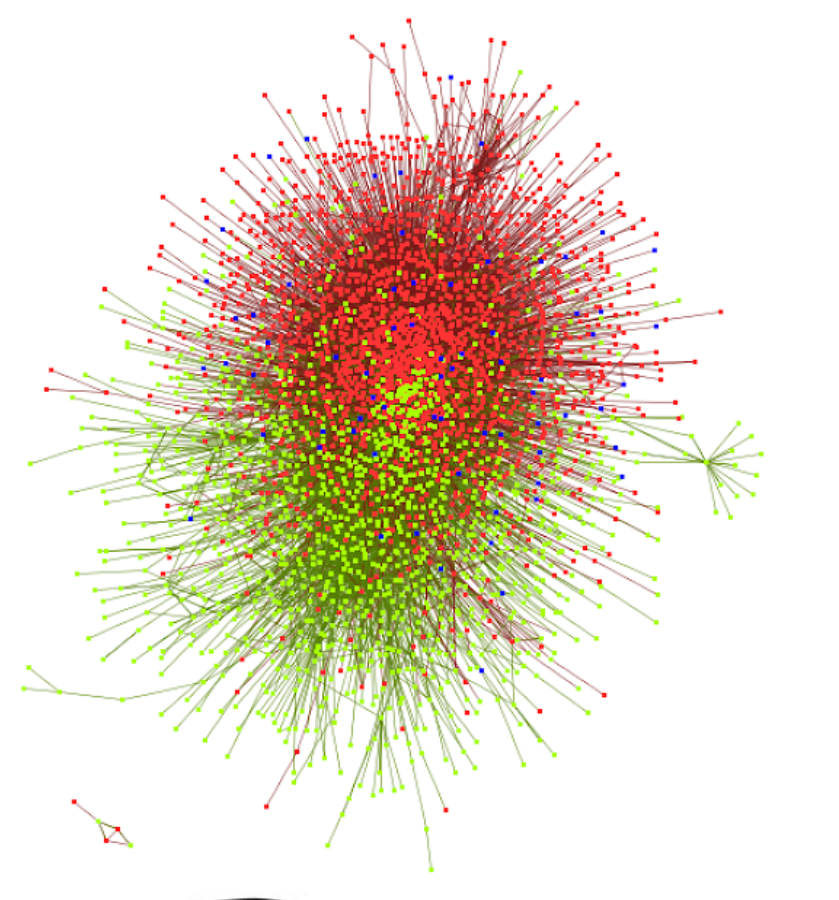}
    \label{fig:mention}
\end{figure}
Along with the linguistic analysis, we also compute various network level measures on the communication networks of the two target groups. These measures include the \emph{network density}, \emph{EI index}, and \emph{echo-chamberness}. We also visualize the three communication networks as shown in figure \ref{fig:mention}. All the network-based measures, and graphs were computed using ORA-PRO \cite{altman2018ora,carley2017ora}. 

We observe that anti-vaccination communities tend to have higher network density, negative EI indices with higher absolute values, and higher echo-chamberness across all the communication networks. On the other hand, the EI index for the pro-vaccination communities is positive for mention and retweet networks displaying dominance of external ties. A summary of network-level measures can be found in Table \ref{tab:networkanalysis}. Interestingly from the network graphs we can observe that on some level the two competing groups are almost detached. This is specifically visible in the retweet network graph in Figure \ref{fig:mention}.

\begin{table}[!ht]
\small
\caption{This table shows the network-level measures for the three types of networks: mention network, retweet network, and reply network}
\centering
\begin{tabular}{|c|c|c|c|} 
\hline
Measure & Mention Network & Retweet Network & Reply Network\\
\hline
\hline
Network Density & 1.7e-5 & 1.1e-5  & 3.1e-6\\ 
Network Density (Pro) & 1.5e-5 & 1.0e-5 & 2.2e-6\\ 
Network Density (Anti) & 4.1e-5 & 3.2e-5 & 6.3e-6\\
\hline
EI Index (Pro) & 0.025 & 0.023 & -0.167\\ 
EI Index (Anti) & -0.276 & -0.432 & -0.572\\
\hline
Echo-chamberness (Pro) & 0.0064334823 & 0.005364444 & 0.0043579605\\ 
Echo-chamberness (Anti) & 0.009268834 & 0.007850341 & 0.005905038\\ 
\hline
\end{tabular}
\vspace{1em}
\label{tab:networkanalysis}
\end{table}

\section{Limitations and Future work}
One minor limitation of our study is that in the data collection phase, the number of collected hashtags for the two communities was unbalanced. This could potentially have introduced some bias in our downstream tasks such as label propagation. A possible limitation pertaining to the network analysis is that we do not normalize our EI indices to avoid losing precision. This, however gives us stronger results as while the nodes in the anti-vaccination network are lower than the pro-vaccination network, the EI index for anti-vaxxers is more negative than pro-vaxxers. Finally, all our analyses are correlational in nature, and do not depict causation. This remains to be one of the important future directions to test whether a certain network characteristic causes linguistic changes in the network or vice-versa.

\section{Conclusion}
In this paper, we have carried out a comparison between two online competing vaccination communities: \emph{pro-vaxxers} and \emph{anti-vaxxers}. We have studied these communities in relation to their linguistic and social interactions. We conduct two kinds of analyses: (i) linguistic, and (ii) network-level. We observe anti-vaxxers to display more frequent usage of linguistic intensification, pronouns, and uncertainty words. We also observe significant differences in the network structures of the two communities with \emph{anti-vaxxers} displaying higher echo-chamberness. These results suggest that anti-vaxxers form a tighter community prone to the presentations of anecdotes, and so may be more resistant to factual knowledge from outside the group.

\section*{Acknowledgement}
This work was partially supported by a fellowship from Carnegie Mellon University’s Center for Machine Learning and Health to Shahan A. Memon. We also acknowledge Lori Levin (LTI,CMU), Bhiksha Raj (LTI,CMU), Rita Singh (LTI,CMU), Matthew Babcock (ISR,CMU), Ingmar Weber (QCRI, HBKU), and members of CMU's Center for Computational Analysis of Social and Organizational Systems (CASOS) for insightful comments and discussions.
%
%
%
%
\bibliographystyle{splncs04}
\bibliography{refs}

\begin{thebibliography}{10}
\providecommand{\url}[1]{\texttt{#1}}
\providecommand{\urlprefix}{URL }
\providecommand{\doi}[1]{https://doi.org/#1}

\bibitem{altman2018ora}
Altman, N., Carley, K.M., Reminga, J.: Ora user’s guide 2018. Carnegie-Mellon
  Univ. Pittsburgh PA Inst of Software Research International, Tech. Rep.
  (2018)

\bibitem{boser2018mothers}
Boser, B.L.: Mothers’ anti-vax to pro-vax conversions. Recovering Argument
  p.~21 (2018)

\bibitem{bradac1995men}
Bradac, J.J., Mulac, A., Thompson, S.A.: Men's and women's use of intensifiers
  and hedges in problem-solving interaction: Molar and molecular analyses.
  Research on Language and Social Interaction  \textbf{28}(2),  93--116 (1995)

\bibitem{broniatowski2016effective}
Broniatowski, D.A., Hilyard, K.M., Dredze, M.: Effective vaccine communication
  during the disneyland measles outbreak. Vaccine  \textbf{34}(28),  3225--3228
  (2016)

\bibitem{broniatowski2018weaponized}
Broniatowski, D.A., Jamison, A.M., Qi, S., AlKulaib, L., Chen, T., Benton, A.,
  Quinn, S.C., Dredze, M.: Weaponized health communication: Twitter bots and
  russian trolls amplify the vaccine debate. American journal of public health
  \textbf{108}(10),  1378--1384 (2018)

\bibitem{brown1992survivorship}
Brown, S.J., Goetzmann, W., Ibbotson, R.G., Ross, S.A.: Survivorship bias in
  performance studies. The Review of Financial Studies  \textbf{5}(4),
  553--580 (1992)

\bibitem{bryden2013word}
Bryden, J., Funk, S., Jansen, V.A.: Word usage mirrors community structure in
  the online social network twitter. EPJ Data Science  \textbf{2}(1), ~3 (2013)

\bibitem{carley2017ora}
Carley, K.M.: Ora: A toolkit for dynamic network analysis and visualization.
  (2017)

\bibitem{dredze2017vaccine}
Dredze, M., Wood-Doughty, Z., Quinn, S.C., Broniatowski, D.A.: Vaccine
  opponents' use of twitter during the 2016 us presidential election:
  Implications for practice and policy. Vaccine  \textbf{35}(36),  4670--4672
  (2017)

\bibitem{duseja2019sociolinguistic}
Duseja, N., Jhamtani, H.: A sociolinguistic study of online echo chambers on
  twitter. In: Proceedings of the Third Workshop on Natural Language Processing
  and Computational Social Science. pp. 78--83 (2019)

\bibitem{evans2016stance}
Evans, A.: Stance and identity in twitter hashtags. Language@ internet
  \textbf{13}(1) (2016)

\bibitem{gibbons2018pronouns}
Gibbons, A., Macrae, A.: Pronouns in literature: Positions and perspectives in
  language. Springer (2018)

\bibitem{giuffre2015cultural}
Giuffre, K.: Cultural production in networks  (2015)

\bibitem{hosman1989evaluative}
Hosman, L.A.: The evaluative consequences of hedges, hesitations, and
  intensifies: Powerful and powerless speech styles. Human communication
  research  \textbf{15}(3),  383--406 (1989)

\bibitem{johnson2020online}
Johnson, N.F., Vel{\'a}squez, N., Restrepo, N.J., Leahy, R., Gabriel, N.,
  El~Oud, S., Zheng, M., Manrique, P., Wuchty, S., Lupu, Y.: The online
  competition between pro-and anti-vaccination views. Nature pp.~1--4 (2020)

\bibitem{kim2014sociolinguistic}
Kim, S., Weber, I., Wei, L., Oh, A.: Sociolinguistic analysis of twitter in
  multilingual societies. In: Proceedings of the 25th ACM conference on
  Hypertext and social media. pp. 243--248 (2014)

\bibitem{krackhardt1988informal}
Krackhardt, D., Stern, R.N.: Informal networks and organizational crises: An
  experimental simulation. Social psychology quarterly pp. 123--140 (1988)

\bibitem{usnews2}
Levy, G.: Public confidence in vaccines sags, new report finds.
  url={https://www.usnews.com/news/health-care-news/articles/2018-05-21/public-confidence-in-vaccines-sags-new-report-finds}

\bibitem{nerbonne2014secret}
Nerbonne, J.: The secret life of pronouns. what our words say about us.
  Literary and Linguistic Computing  \textbf{29}(1),  139--142 (2014)

\bibitem{sanawi2017vaccination}
Sanawi, J.B., Samani, M.C., Taibi, M.: \# vaccination: Identifying influencers
  in the vaccination discussion on twitter through social network
  visualisation. International Journal of Business and Society
  \textbf{18}(S4),  718--726 (2017)

\bibitem{smelser2001international}
Smelser, N.J., Baltes, P.B., et~al.: International encyclopedia of the social
  \& behavioral sciences, vol.~11. Elsevier Amsterdam (2001)

\bibitem{tyagi2020brims}
Tyagi, A., Babcock, M., Carley, K.M., Sicker, D.C.: Polarizing tweets on
  climate change. To appear in International Conference SBP-BRiMS  (2020)

\bibitem{tyagi2020computational}
Tyagi, A., Field, A., Lathwal, P., Tsvetkov, Y., Carley, K.M.: A computational
  analysis of polarization on indian and pakistani social media (2020)

\bibitem{wasserman1994social}
Wasserman, S., Faust, K., et~al.: Social network analysis: Methods and
  applications, vol.~8. Cambridge university press (1994)

\bibitem{xiaojin2002learning}
Xiaojin, Z., Zoubin, G.: Learning from labeled and unlabeled data with label
  propagation. Tech. Rep., Technical Report CMU-CALD-02--107, Carnegie Mellon
  University  (2002)

\bibitem{young2004systemic}
Young, L., Harrison, C.: Systemic functional linguistics and critical discourse
  analysis: Studies in social change. A\&C Black (2004)

\end{thebibliography}




\end{document}